\documentclass[final,journal,10pt,letterpaper]{IEEEtran}
\usepackage[pdftex]{graphicx}
\usepackage[cmex10]{amsmath}
\usepackage[tight,footnotesize]{subfigure}
\usepackage{url}
\usepackage{amsfonts}
\usepackage{amssymb}
\usepackage{bm}
\usepackage{array}
\usepackage{multicol}
\usepackage{cite}
\usepackage{url}
\usepackage[colorlinks, bookmarks=false, citecolor=blue, linkcolor=red, urlcolor=blue]{hyperref}
\DeclareMathOperator*{\argmin}{arg\,min}

\begin{document}

\title{Resistance distance criterion \\ for optimal slack bus selection}
\author{Tommaso~Coletta,
        and Philippe~Jacquod,~\IEEEmembership{Member,~IEEE}
\thanks{T. Coletta and P. Jacquod are with the School of Engineering of the University of Applied Sciences of Western Switzerland
CH-1951 Sion, Switzerland. Emails: (tommaso.coletta, philippe.jacquod)@hevs.ch.}
}

\markboth{THIS VERSION: JULY 07, 2017}%
{Coletta \MakeLowercase{\textit{et al.}}: Resistance distance criterion for optimal slack bus selection}

\maketitle

\begin{abstract}
We investigate the dependence of transmission losses on the choice of a slack bus
in high voltage AC transmission networks.
We formulate a transmission loss minimization problem in terms of slack variables 
representing the additional power injection that each generator provides to compensate 
the transmission losses.
We show analytically that for transmission lines having small, homogeneous resistance over reactance ratios $\bm{r/x\ll1}$, 
transmission losses are generically minimal in the case of a unique \textit{slack bus} instead of a distributed slack bus.
For the unique slack bus scenario, to lowest order in $\bm{r/x}$, transmission losses depend linearly on a resistance 
distance based indicator measuring the separation of the slack bus candidate from the rest of the network.
We confirm these results numerically for several IEEE and Pegase testcases, and show that our predictions 
qualitatively hold also in the case of lines having inhomogeneous $\bm{r/x}$ ratios, 
with optimal slack bus choices reducing transmission losses by $\bm{10}\bm{\%}$ typically.
\end{abstract}
\begin{IEEEkeywords}
Resistance distance, power flow equations, distributed slack bus, transmission losses,
participation factors.
\end{IEEEkeywords}


\section{Introduction}

\IEEEPARstart{T}{he} power flow problem relates the injected and consumed power at every bus of an AC electric network
to the power transmitted and dissipated along the branches of the network.
The dissipative character of transmission lines implies that the perfect balance between 
power injection and consumption is not realized a priori because
transmission losses are a function of the operational state of the system.
To overcome this aspect and solve the power flow problem two standard approaches exist: 
i) the DC approximation, and ii) the slack bus treatment of the full AC problem.

Several variants of the DC approximation exist \cite{Bialek08,Grainger94}, 
but all involve a linearization of the power flow problem.
Once the contribution of shunt elements is incorporated into an effective injected/consumed power, 
the DC approximation automatically enforces the perfect balance between consumed and injected power.
In this framework, transmission losses can only be estimated. 
This can be done either starting from a known solution to the full AC problem as 
in the $\alpha-$matching model \cite{Stott09,Qi12} or through iterative procedures requiring to solve a sequence of DC 
power flow problems updating the power injections at each step \cite{Simpson16}.

In contrast, the slack bus approach is used to tackle the full AC power flow problem.
This standard textbook procedure requires to promote one of the generators of the network
to be the voltage magnitude and phase reference of the system \cite{Bialek08,Grainger94}. 
This involves leaving undetermined the active and reactive power injections at this specific generator
commonly called the \textit{slack} or \textit{swing bus}. 
These quantities are determined by solving numerically the power flow problem and 
account for the power imbalance necessary to compensate transmission losses.
Standard heuristic criteria for suitable slack bus selection include: i) sufficiently large available production to
compensate the power imbalance, ii) strong network connectivity, and iii)
a bus voltage that leads all other voltages in the network \cite{Freris68}.

Additionally to these heuristic criteria, it is desirable to pick the slack bus through an algorithmic approach.
Some pioneering work in this direction investigated the influence of the slack bus choice on the convergence of the 
numerical methods used to solve the AC power flow problem \cite{Freris68}.
More recent works investigated the aspect of slack bus generation constraints in the case of fluctuating nodal powers
\cite{Dimitrovski04} or, of direct relevance to this work, the slack bus choice that minimizes the power imbalance 
\cite{Exposito04}. 
In parallel, distributed slack bus approaches have also been developed as alternatives to a unique slack bus.
The additional power injection necessary to satisfy the power balance is shared among
several generators, the contribution of each generator being encoded in a vector of participation factors 
\cite{Meisel93,Guoyu85}.
In this formulation the single slack bus case consists of a particular participation vector.

Ref. \cite{Exposito04} formulated a transmission loss minimization problem in terms of the slack variables and 
suggested that for positive participation factors the single slack bus scenario is generically the optimal solution.
More generally one may want simple and computationally inexpensive criteria to determine which elements of the network 
-- nodes or lines -- are the most critical with respect to one specific objective.
In the present case the specific objective is minimizing the transmission losses, but other 
cases include: optimal virtual inertia allocation \cite{Poola17}, or determining critical nodes where faults 
affect network operation most strongly \cite{Milanovic17}.
Following this direction, we revisit from a graph theoretical perspective the problem addressed in \cite{Exposito04}
and propose a resistance distance \cite{Stephenson89, Klein93} based indicator to determine the optimal generator, or generators which
minimize transmission losses.

Our strategy is to start from the lossless power flow problem \cite{Dorfler13}, 
for which a solution is assumed to be known a priori,  
and treat dissipative effects as a perturbation around that solution.
This approach is justified for high voltage AC transmission lines which are characterized by 
small $r/x$ ratios (admittance dominated by its imaginary part)
and therefore weak transmission losses. In the spirit of Ref.~\cite{Exposito04}, to account for the power imbalance resulting from 
transmission losses, we introduce a 
vector of slack variables instead of a single slack bus, 
and formulate a transmission loss minimization problem in terms of these slack variables.
We show analytically that, to leading order in $r/x$, transmission losses are generically minimized by choosing a 
single slack bus. 
Furthermore, we find a simple graph theoretical indicator based on the resistance distance \cite{Stephenson89, Klein93}, 
which is computed from the solution of the lossless power flow problem only.
It easily allows to determine the optimal slack bus choice from a transmission losses point of view.

We confirm our analytical predictions by performing numerical investigations on several IEEE and Pegase testcases 
\cite{Fliscounakis13, Josz16}.
Our numerics indicate that an optimal slack bus choice can reduce the total transmission losses by $10\%$, and that the
tabulated slack bus generators of several testcases are not always the optimal ones.
Our work further complements the results of 
Ref.~\cite{Exposito04} providing an intuitive and computationally inexpensive graph theoretical indicator to interpret and predict the optimal slack 
bus choice.
Our indicator, specifying the generators which are most relevant for transmission losses, could be used to provide a hot start 
to more sophisticated optimal power flow algorithms with the advantage of reducing the dimension of the parameter space to be investigated,
thereby reducing the computational effort.

This paper is organized as follows: Section \ref{sec:Resistance distance} 
recalls the definition of the resistance distance metric. Section \ref{sec:Losses estimate} presents
our leading estimate of the transmission losses in the AC power flow problem in the general case 
of a distributed slack bus. Section \ref{sec:Single generation compensation} relates the transmission losses 
to the graph theoretical notion of resistance distance. Sections \ref{sec:Multiple generation compensation} 
and \ref{sec:NLO} address the transmission loss minimization problem.
A brief conclusion is given in Section \ref{sec:Conclusion}. 

\section{Resistance distance}\label{sec:Resistance distance}
Let ${\bm L}$ be the Laplacian of a weighted undirected graph $\mathcal{G}$ composed of $N$ nodes and $|\mathcal{E}|$ edges.
We denote by $w_{ij}=w_{ji}\geq0$ the weight of the edge connecting nodes $i$ and $j$ while
$\{\lambda_1,\lambda_2,\ldots, \lambda_N\}$ and $\{{\bm u}^{(1)},\ldots,{\bm u}^{(N)}\}$ 
are the eigenvalues and eigenvectors of the weighted Laplacian respectively.
The Laplacian has one eigenvalue equal to zero, $\lambda_1=0$, with the corresponding eigenvector $\bm u^{(1)}=(1,\ldots,1)/\sqrt{N}$. 
Given the matrix ${\bm \Gamma}$ defined as
\begin{equation}
 {\bm \Gamma}={\bm L} + \frac{1}{N}{\bm u^{(1)}}^\top \bm u^{(1)}\,,
\end{equation}
the resistance distance $\Omega_{ij}$ between nodes $i$ and $j$ is defined as \cite{Stephenson89,Klein93}
\begin{equation}\label{eq:Resistance distance}
 \Omega_{ij}=\Gamma_{ii}^{-1}+\Gamma_{jj}^{-1}-2\Gamma_{ij}^{-1}\,.
\end{equation}
By construction, $\bm L$ and $\bm \Gamma$ share the same eigenvectors, and the eigenvalues of 
$\bm \Gamma$ are $\{1/N, \lambda_2,\ldots,\lambda_N\}$. Expressing 
$\bm \Gamma^{-1}$ as a function of the eigenvectors $\bm u^{(i)}$ and eigenvalues $\lambda_i$
one has
\begin{align}\label{eq:Gamma inverse}
\nonumber
  \Gamma_{ij}^{-1} & =  N u_i^{(1)}u_j^{(1)} + \sum_{l\geq2}\lambda_l^{-1}u_i^{(l)}u_j^{(l)} \\ 
                   & = 1 + (L^{-1})_{ij}\,,
\end{align}
where ${\bm L}^{-1}$ is the Moore-Penrose pseudoinverse of $\bm{L}$ defined as 
$\bm{L}^{-1}={\bm T}\textrm{diag}(\{0,\lambda_2^{-1},\ldots,\lambda_N^{-1}\}){\bm T}^\top$,
and ${\bm T}$ is the matrix with ${\bm u}^{(i)}$ as its $i^\textrm{th}$ column.
Injecting back Eq.~(\ref{eq:Gamma inverse}) into Eq.~(\ref{eq:Resistance distance}) leads to the following 
equivalent definition of the resistance distance
\begin{equation}\label{eq:Resistance distance compact}
 \Omega_{ij}=\sum_{l\geq 2}\lambda_l^{-1}\left(u^{(l)}_i-u^{(l)}_j\right)^2\,.
\end{equation}

The graph theoretical metric $\Omega_{ij}$ is a distance in the mathematical sense
since $\Omega_{ij}\geq0$, $\Omega_{ij}=\Omega_{ji}$, and $\Omega_{ij}\leq\Omega_{il}+\Omega_{lj}\,\forall\, i,j,l$.
Furthermore it is known as the \textit{resistance} distance because 
if one replaces the edges of $\mathcal{G}$ by resistors with a conductance $1/R_{ij}\equiv w_{ij}$, 
then $\Omega_{ij}$
is equal to the equivalent network resistance when a current is injected at node $i$ and extracted at node $j$
with no injection anywhere else.
Accordingly $\Omega_{ij}$ accounts for the contributions of all the parallel paths between $i$ and $j$, as it should.
The existence of multiple parallel paths between two nodes reduces the resistance distance between them. 

\section{Estimating transmission losses in high voltage AC networks}\label{sec:Losses estimate}
We model AC electric networks by a set of complex voltages $V_ie^{i\theta_i}$ 
and a set of active power injections, $P_i>0$, or consumptions, $P_i<0$, defined at every node of a graph $\mathcal{G}$.
The $|\mathcal{E}|$ edges of $\mathcal{G}$ represent the electrical connections between the different buses of the network.
The transmission lines connecting any two nodes $i$ and $j$ are characterized by the real and imaginary parts of the admittance,
i.e. the conductance $g_{ij}$ and the susceptance $b_{ij}$ respectively.
We assume $g_{ij}/b_{ij}=\gamma$ for every line in the network. 
This amounts to considering that all lines are made of the same material
and have the same geometrical proportions.
In the case of high voltage AC transmission networks, lines are mostly susceptive and typically $\gamma\leq 0.2$.
 
Given a set of power injections and voltage magnitudes, we consider solutions $\{\theta_i^{(0)}\}$
to the lossless power flow equations \cite{Dorfler13}
\begin{equation}\label{eq:Power flow lossless}
 P_i=\sum_{j\sim i}b_{ij}V_iV_j\sin(\theta_i^{(0)}-\theta_j^{(0)})\,, \quad \forall i\,,
\end{equation}
where $j\sim i$ indicates all nodes $j$ connected to node $i$.

In the lossless line approximation the power balance between injection and consumption 
is satisfied, $\sum_iP_i=0$.
We express Eq.~(\ref{eq:Power flow lossless}) in vector form in terms of the vector of power 
injections ${\bm P}=(P_1,\ldots,P_N)$ as 
\begin{equation}\label{eq:Power flow lossless compact}
 {\bm P}={\bm B}{\bm\Lambda}\cdot\bm{\mathrm{sin}}({\bm B}^\top{\bm \theta}^{(0)}) 
\end{equation}
where $\bm{\mathrm{sin}}({\bm x}) \equiv (\sin(x_1 ), \ldots , \sin(x_{|\mathcal{E}|}))$
for ${\bm x} \in \rm I\!R^{|\mathcal{E}|}$, ${\bm \Lambda} \in \rm I\!R^{|\mathcal{E}|\times|\mathcal{E}|}$ is the diagonal matrix of edge
weights ${\bm \Lambda}=\textrm{diag}(\{b_{ij}V_iV_j\})$, and $\bm B\in\rm I\!R^{N\times|\mathcal{E}|}$ is the incidence matrix of the graph defined as
\begin{equation}
\begin{array}{ccc}
 {\bm B}_{il}=\left\{
 \begin{array}{cl}
  1 \quad &\textrm{if }i\textrm{ is the source of edge }l \,,\\
  -1 \quad &\textrm{if }i\textrm{ is the sink of edge }l \,,\\
   0 \quad &\textrm{otherwise} \,.\\
  \end{array}
 \right.
\end{array}
\end{equation}

In the case of dissipating lines, power injections and consumptions are modified with respect to the lossless line approximation. 
The power flow equations read
\begin{equation}\label{eq:Power flow losses}
P_i+\delta P_i = V_i\sum_{j\sim i}b_{ij}V_j\sin(\theta_i-\theta_j)+g_{ij}\left[V_i-V_j\cos(\theta_i-\theta_j)\right].
\end{equation}
Our approach, consists in introducing a vector of slack variables $\bm{\delta P}=(\delta P_1,\ldots,\delta P_N)$. 
instead of a single slack variable $\delta P_g$,
similarly to Ref.~\cite{Exposito04}.
The total dissipated active power $D$ is obtained by summing Eq.~(\ref{eq:Power flow losses}) over all nodes.
This yields
\begin{equation}\label{eq:Total dissipation}
 D=\sum_i\delta P_i = \sum_{\langle i,j\rangle}g_{ij}\left[V_i^2+V_j^2-2V_iV_j\cos(\theta_i-\theta_j)\right]\,,
\end{equation}
where $\langle i,j\rangle$ indicates that the sum runs over all edges of the network.
Eq.~(\ref{eq:Total dissipation}) implies that a priori all nodes can contribute to compensate the losses in the network.
This differs from the standard textbook method using a single slack bus \cite{Bialek08,Grainger94}.

Given that the full AC power flow problem, Eq.~(\ref{eq:Power flow losses}), consists of a linear combination of analytic functions 
of the voltage phases, its solutions are expected to be analytic functions of the parameter $\gamma$ and may be formally 
written as 
\begin{equation}\label{eq:Analytic solutions}
 \bm{\theta}(\gamma) = \bm{\theta}^{(0)} + \sum_{n\geq1}\gamma^n\bm{\delta\theta}^{(n)}\,, \quad 
 \bm{\delta P}(\gamma) = \sum_{n\geq1}\gamma^n\bm{\delta P}^{(n)}\,,
\end{equation}
where one recovers the lossless solution $\bm{\theta}(0)=\bm{\theta}^{(0)}$, $\bm{\delta P}(0)=0$ when $\gamma\rightarrow0$.
The representation of the solutions to the power flow problem given in Eq.~(\ref{eq:Analytic solutions})
allows to perform systematic expansions of $D$ in powers of $\gamma$.
Since $\gamma\ll1$ in high voltage AC electrical networks, 
we treat conductance contributions perturbatively with respect to the lossless case and truncate the resulting expansion 
to lower orders in $\gamma$.

Up to, and including order $\gamma^2$, Eq.~(\ref{eq:Total dissipation}) becomes
\begin{align}\label{eq:Dissipation expansion}
\nonumber
D = &\, \gamma\sum_{\langle i,j\rangle} b_{ij}\left[V_i^2+V_j^2-2V_iV_j\cos(\theta_i^{(0)}-\theta_j^{(0)})\right] \\
\nonumber
  &+\gamma^2\sum_{\langle i,j\rangle}2b_{ij}V_iV_j\sin(\theta_i^{(0)}-\theta_j^{(0)})(\delta\theta_i^{(1)}-\delta\theta_j^{(1)})\\
  &+\mathcal{O}(\gamma^3)\,.
\end{align}
The first term on the right-hand side of Eq.~(\ref{eq:Dissipation expansion}) 
indicates that the leading contribution to the total 
dissipation is independent of $\bm{\delta P}$,  
\begin{equation}\label{eq:D0}
 D^{(0)}\equiv\gamma\sum_{\langle i,j\rangle}b_{ij}\left[V_i^2+V_j^2-2V_iV_j\cos(\theta_i^{(0)}-\theta_j^{(0)})\right]\,.
\end{equation} 
This quantity is linear in $\gamma$. It corresponds to the dissipation one would have if the 
voltage phases were unaffected by ohmic dissipation in the transmission lines.

The voltage phases are modified in the presence of dissipation and depend explicitly on $\bm{\delta P}$.
Next, we compute the leading order corrections to the phases $\bm{\delta \theta}^{(1)}$ and relate them to the power dispatch $\bm{\delta P}$. 
Inserting Eq.~(\ref{eq:Analytic solutions}) into Eq.~(\ref{eq:Power flow losses}) and expanding both sides of the equation to first order in $\gamma$ we obtain
\begin{align}
\nonumber
 \delta P_i^{(1)} =&\sum_{j\sim i}b_{ij}V_iV_j\cos(\theta_i^{(0)}-\theta_j^{(0)})(\delta\theta_i^{(1)}-\delta\theta_j^{(1)})\\
             &+\sum_{j\sim i}b_{ij}V_i\left[V_i-V_j\cos(\theta_i^{(0)}-\theta_j^{(0)})\right] \quad \forall i\,.
\end{align}
This can be expressed in vector form as 
\begin{equation}\label{eq:Equation delta theta}
 \bm{\delta P}^{(1)} = {\bm v} + {\bm L}\bm{\delta\theta}^{(1)}\,,
\end{equation}
where
$v_i=\displaystyle\sum_{j\sim i}b_{ij}V_i\left[V_i-V_j\cos(\theta_{i}^{(0)}-\theta_j^{(0)})\right]$, 
and ${\bm L}$ is the weighted Laplacian
\begin{equation}\label{eq:Laplacian}
 L_{ij}=\left\{\begin{array}{cl}
               \displaystyle -b_{ij}V_iV_j\cos{(\theta_i^{(0)}-\theta_j^{(0)})} &\quad j \sim i\,,  \\[3mm]
               \displaystyle \sum_{l\sim i}b_{il}V_iV_l\cos{(\theta_i^{(0)}-\theta_l^{(0)})} &\quad i=j\,, \\[3mm]
               \displaystyle 0 &\quad \textrm{otherwise}\,.
               \end{array}\right.
\end{equation}
Using the eigenvalues and eigenvectors of ${\bm L}$,  $\{\lambda_1=0,\lambda_2,\ldots,\lambda_N\}$
and $\{{\bm u}^{(1)},{\bm u}^{(2)},\ldots,{\bm u}^{(N)}\}$, 
and ${\bm L}^{-1}{\bm L}=\mathbb{I}-{\bm u}^{(1)}{{\bm u}^{(1)}}^\top$, we invert Eq.~(\ref{eq:Equation delta theta}). 
We obtain
\begin{align}\label{eq:Delta theta}
 \delta\theta_i^{(1)}-\delta\theta_j^{(1)}=\sum_{l\geq2}\lambda_l^{-1}\left(u_i^{(l)}-u_j^{(l)}\right)\left[{{\bm u}^{(l)}}^\top\cdot(\bm{\delta P}^{(1)}-{\bm v})\right]\,.
\end{align}
Injecting this expression back into Eq.~(\ref{eq:Dissipation expansion}) gives
\begin{multline}\label{eq:Dissipation expansion2}
 D = D^{(0)} +\gamma^2\sum_{\substack{\langle i,j\rangle \\ l\geq2}}2b_{ij}V_iV_j\sin(\theta_i^{(0)}-\theta_j^{(0)})\lambda_l^{-1}\left(u_i^{(l)}-u_j^{(l)}\right) \\
 \cdot\left[{{\bm u}^{(l)}}^\top\cdot(\bm{\delta P}^{(1)}-{\bm v})\right] +\mathcal{O}(\gamma^3)\,.
\end{multline}
We further simplify this expression using
\begin{align}\label{eq:Identity}
\nonumber
{{\bm u}^{(l)}}^\top \bm P =&\left({\bm B}^\top{\bm u}^{(l)}\right)^\top{\bm \Lambda}  \cdot\bm{\mathrm{sin}}\left({\bm B}^T\bm\theta^{(0)}\right) \\
                        =&\sum_{\langle i,j\rangle}b_{ij}V_iV_j\left(u_i^{(l)}-u_j^{(l)}\right)\sin\left(\theta^{(0)}_i-\theta^{(0)}_j\right)\,,
\end{align}
which follows from Eq.~(\ref{eq:Power flow lossless compact}).
The total dissipation up to and including order $\mathcal{O}(\gamma^2)$ is given by
\begin{equation}\label{eq:Total dissipation compact}
 D \approx D^{(0)}+2\gamma^2\sum_{l\geq2}\lambda_l^{-1}\left({{\bm u}^{(l)}}^\top\cdot \bm P\right)
 \left[{{\bm u}^{(l)}}^\top\cdot(\bm{\delta P}^{(1)}-{\bm v})\right]\,.
\end{equation}

We finally isolate the leading contribution depending
on $\bm{\delta P}$
\begin{equation}\label{eq:Total dissipation dispatch}
D|_{\bm{\delta P}^{(1)}}=2\gamma^2\sum_{l\geq2}\lambda_l^{-1}\left({{\bm u}^{(l)}}^\top\cdot \bm P\right)
\left({{\bm u}^{(l)}}^\top\cdot\bm{\delta P^{(1)}}\right)\,.
\end{equation}
To the best of our knowledge, the expression given in Eq.~(\ref{eq:Total dissipation dispatch}) is new. 
It emphasizes that different choices of $\bm{\delta P}$ lead to different amounts of dissipation. 
In the next two paragraphs we discuss particular
choices of $\bm{\delta P}^{(1)}$ which minimize transmission losses in two different situations.

\subsection{Single generator compensation}\label{sec:Single generation compensation}
We first focus on the case where only one generator (labeled $g$) produces all the additional power necessary to compensate for the transmission losses.
This case is the standard approach, where generator $g$ is the slack bus \cite{Bialek08}.
In this case the components of $\bm{\delta P}$ are given by $\delta P_i=\delta P_g \delta_{i,g}$, and since
the total dissipation is equal to $\sum_i \delta P_i$, to leading order in $\gamma$ one has
$\gamma\delta P_g^{(1)}=D^{(0)}$.
Eq.~(\ref{eq:Total dissipation dispatch}) becomes
\begin{equation}\label{eq:Total dissipation dispatch2}
D|_{\bm{\delta P}^{(1)}}=2\gamma D^{(0)}
                    \sum_{l\geq2}\lambda_l^{-1}\left({{\bm u}^{(l)}}^\top\cdot \bm P\right)u^{(l)}_g\,.
\end{equation}
Next we show how Eq.~(\ref{eq:Total dissipation dispatch2}) can be reformulated in a much more insightful way in terms of the resistance distance.

According to Eq.~(\ref{eq:Resistance distance compact}), the resistance distance between generator $g$ and 
any other node $i$ is
\begin{equation}
 \Omega_{gi}=\sum_{l\geq 2}\lambda_l^{-1}\left(u^{(l)}_i-u^{(l)}_g\right)^2\,.
\end{equation}
Compared to the discussion in Section \ref{sec:Resistance distance}, $\lambda_l$ and $\bm{u}^{(l)}$
are now the eigenvalues and eigenvectors of a weighted Laplacian
with edge weights $b_{ij}V_iV_j\cos(\theta_i^{(0)}-\theta_j^{(0)})$.
Accordingly, the ohmic resistance between two connected nodes is 
$R_{ij}\equiv[b_{ij}V_iV_j\cos(\theta_i^{(0)}-\theta_j^{(0)})]^{-1}$.
This non linear relation between edge resistance and voltage phase difference implies that when two nodes are separated by heavily loaded 
transmission lines the resistance distance separating them is large.

Defining the vector of resistance distances with respect to bus $g$, 
$\bm{\Omega_g}\equiv(\Omega_{g1},\ldots,\Omega_{gN})$ we have
\begin{align}\label{eq:Product resistance distance power}
\nonumber
 \bm{\Omega_g}^\top \cdot \bm{P} &= 
 \displaystyle \sum_i\sum_{l\geq 2}\lambda_l^{-1}P_i\left({u^{(l)}_i}^2-2u^{(l)}_g u^{(l)}_i+{u^{(l)}_g}^2\right) \\
 &=
 \displaystyle \sum_{l\geq 2}\lambda_l^{-1}\left[\sum_iP_i{u^{(l)}_i}^2-2u^{(l)}_g\left({{\bm u}^{(l)}}^\top\cdot \bm P\right)\right]\,,
\end{align}
where we used $\sum_i P_i=0$.
Injecting Eq.~(\ref{eq:Product resistance distance power}) into Eq.~(\ref{eq:Total dissipation dispatch2}),
finally gives
\begin{equation}\label{eq:Total dissipation dispatch3}
D|_{\bm{\delta P}^{(1)}}=\gamma D^{(0)}
                    \left[-\bm{\Omega_g}^\top \cdot \bm{P}+\sum_{l\geq2,i}
                    \lambda_l^{-1}P_i{u_i^{(l)}}^2\,\right]\,.
\end{equation}

To the best of our knowledge Eq.~(\ref{eq:Total dissipation dispatch3}) is a new result.
It shows that the leading order contribution to $D$ depending on the slack bus is proportional to $-\bm{\Omega}_g^\top\cdot\bm P$.
The second term in the right-hand side of Eq.~(\ref{eq:Total dissipation dispatch3}) is always the same 
regardless of the slack bus choice.
Furthermore, $\bm{\Omega_g}^\top \cdot \bm{P}$ can be interpreted as an average of the resistance distance 
between generator $g$ and all the other nodes of the network weighted by the power injections $\bm P$ of the lossless case.
Therefore, determining the optimal choice of slack bus for which total losses will be lowest amounts to 
minimizing $-\bm{\Omega_g}^\top \cdot \bm{P}$ over different generators $g$.
This quantity depends only on the operating conditions of the system in the lossless case
and is therefore easily computed.

The fact that losses depend on $-\bm{\Omega}_g^\top \cdot \bm{P}$ can be understood as follows.
Assume generator $g$ is a large distance away from node $i$ consuming a power $P_i<0$,
then $-\Omega_{gi}P_i$ is a positive quantity which increases losses substantially because the 
extra power to compensate for losses has to travel a long distance from $g$ to $i$. 
Inversely, if node $i$ is a large generator, $P_i>0$, $-\Omega_{gi}P_i$ is negative 
and losses are reduced. This is consistent with the fact that losses in the vicinity of node $i$ are
compensated by the injection at $i$ and not by that of the far away slack bus $g$.

In order to confirm our prediction that losses are linear in $-\bm{\Omega}_g^\top\cdot \bm{P}$, 
we perform numerical simulations 
for the IEEE-57, IEEE-118, Pegase-89, and Pegase-1354 testcases \cite{Fliscounakis13,Josz16}.
We assume that all buses are PV-nodes and neglect shunt susceptances and shunt conductances which, 
for PV-buses, can be absorbed in the active power injection at every node.
First we simulate the lossless case (i.e. $g_{ij}=0$ for all lines) and compute $-{{\bm \Omega}_g}^\top\cdot{\bm P}$ for all 
the large generator buses in the network (i.e. the potential slack bus candidates).
Next we perform a full power flow simulation (including line conductances) each time choosing as slack generator a different candidate. 

Because our analytical result, Eq.~(\ref{eq:Total dissipation dispatch3}), is based on the assumption $\gamma\equiv g_{ij}/b_{ij}$,
we first replace the tabulated values for the line conductances $g_{ij}$ by 
$\gamma b_{ij}$ with $\gamma=0.01,0.1,$ and $0.2$. 
\begin{figure*}[!t]
\centering
  \includegraphics[width=0.98\textwidth]{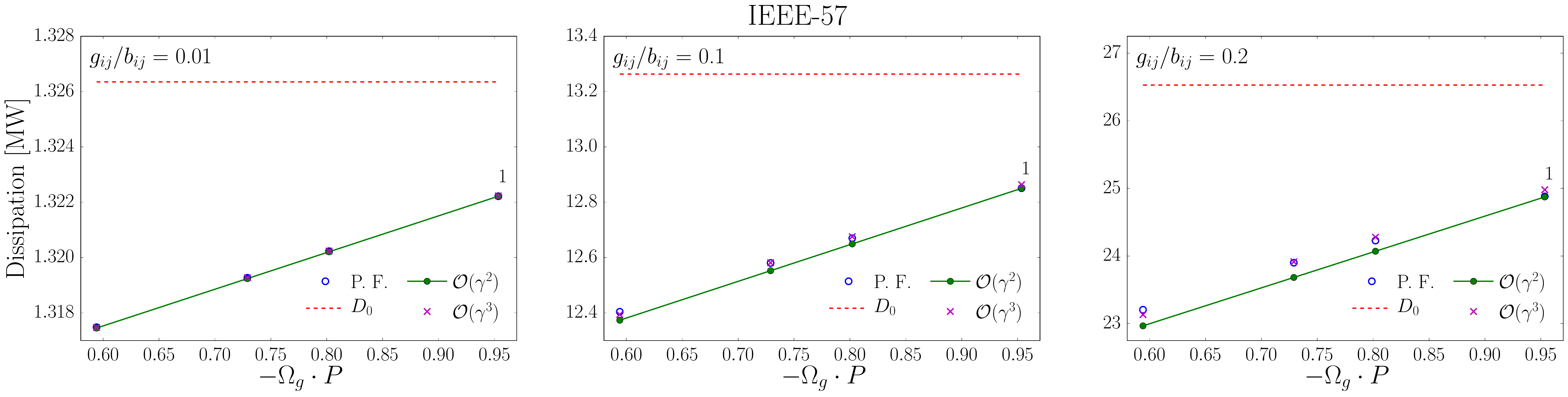}
  \includegraphics[width=0.98\textwidth]{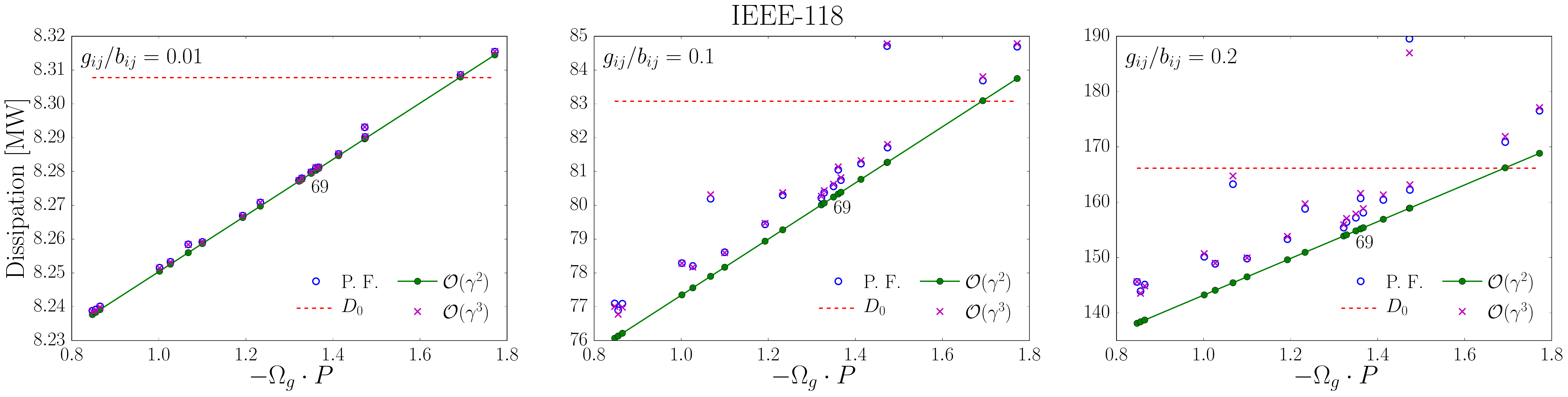}
  \includegraphics[width=0.98\textwidth]{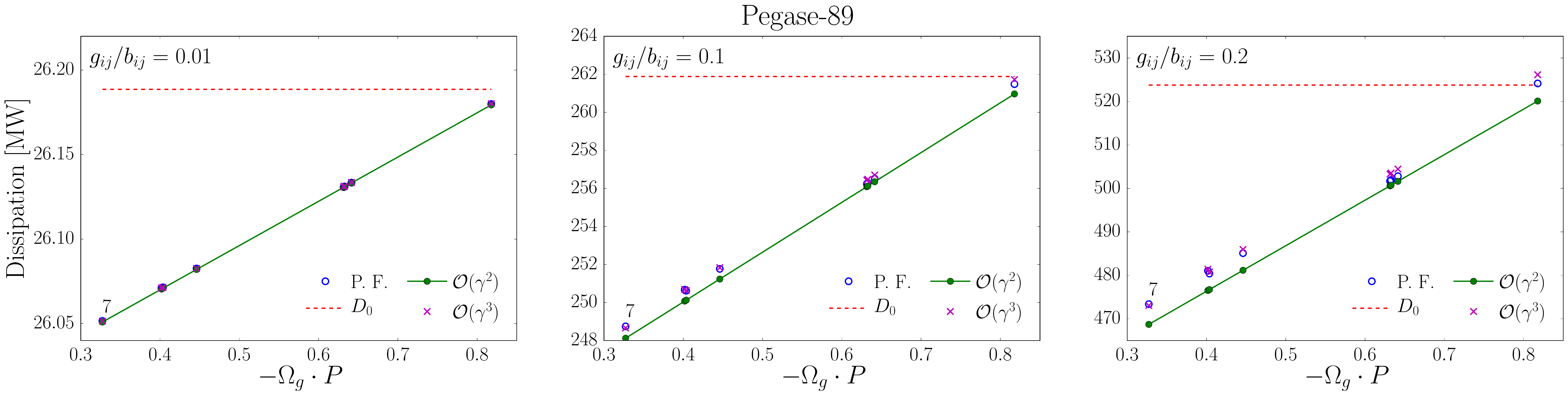}
  \includegraphics[width=0.98\textwidth]{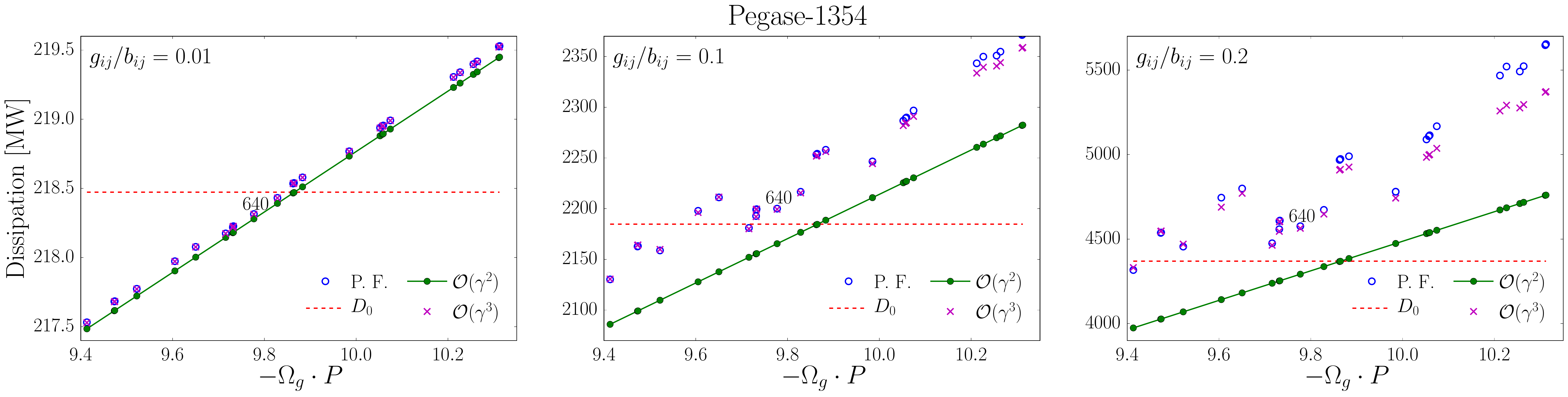}
 \caption{Total dissipation as a function of $-\bm{\Omega_g}^\top \cdot \bm{P}$, 
 for different slack bus choices in the testcases: IEEE-57, IEEE-118, Pegase-89
 and Pegase-1354 \cite{Fliscounakis13, Josz16} (restricted to generators having $P_i>1$GW for the Pegase-1354 testcase).
 Each row corresponds to a different conductance to susceptance ratio $g_{ij}/b_{ij}=\gamma\in\{0.01,0.1,0.2\}$ from left to right. 
 The dashed red line and the green line are $D^{(0)}$ and the expression of Eq.~(\ref{eq:Total dissipation compact}) respectively. 
 Pink crosses are $\mathcal{O}(\gamma^3)$ estimates for the total dissipation [sum of Eq.~(\ref{eq:Total dissipation compact})
 and Eq.~(\ref{eq:Second order term easy})] while blue circles are 
 the Newton Raphson numerical solutions of the full power flow problem.
 The data points annotated by a number indicate the tabulated slack bus generators.}
 \label{fig:testcases uniform lines}
\end{figure*}
Fig.~\ref{fig:testcases uniform lines} shows the total transmission losses
as a function of the weighted resistance distance $-\bm{\Omega}_g^\top\cdot{\bm P}$ for different slack bus choices.
Our results confirm that the leading behavior of the total dissipation as a function of the slack bus choice scales linearly
with the measure $-\bm{\Omega_g}^\top \cdot \bm{P}$ with small deviations appearing as $\gamma$ is increased.
We also see that the slack bus appearing in the testcase dataset 
is not always the choice that minimizes losses.
Total losses differ by up to $30\%$ for different slack bus choices for $\gamma=0.2$ in the Pegase-1354 testcase.

\begin{figure*}[!h]
\centering
  \includegraphics[width=0.49\textwidth]{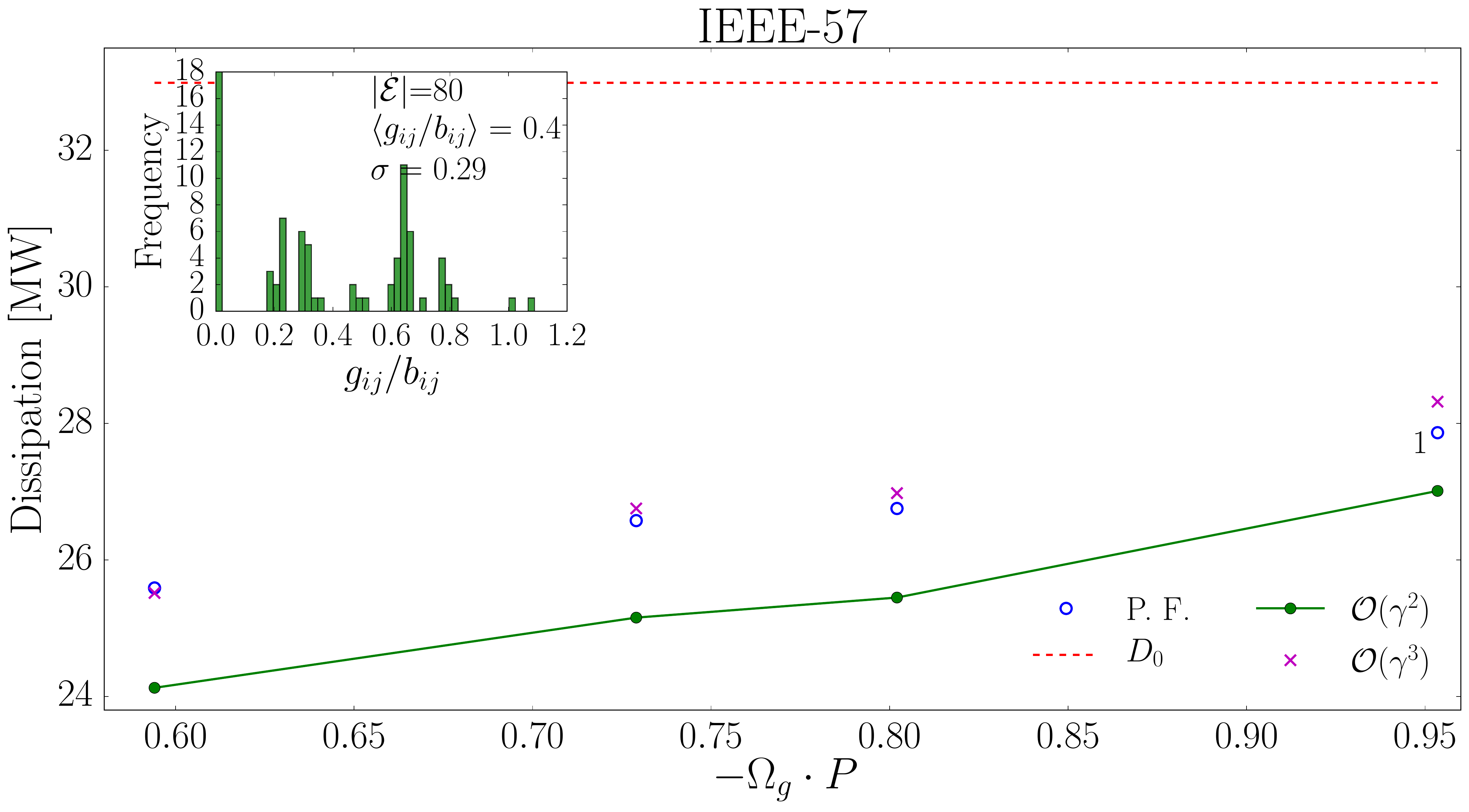}
  \includegraphics[width=0.49\textwidth]{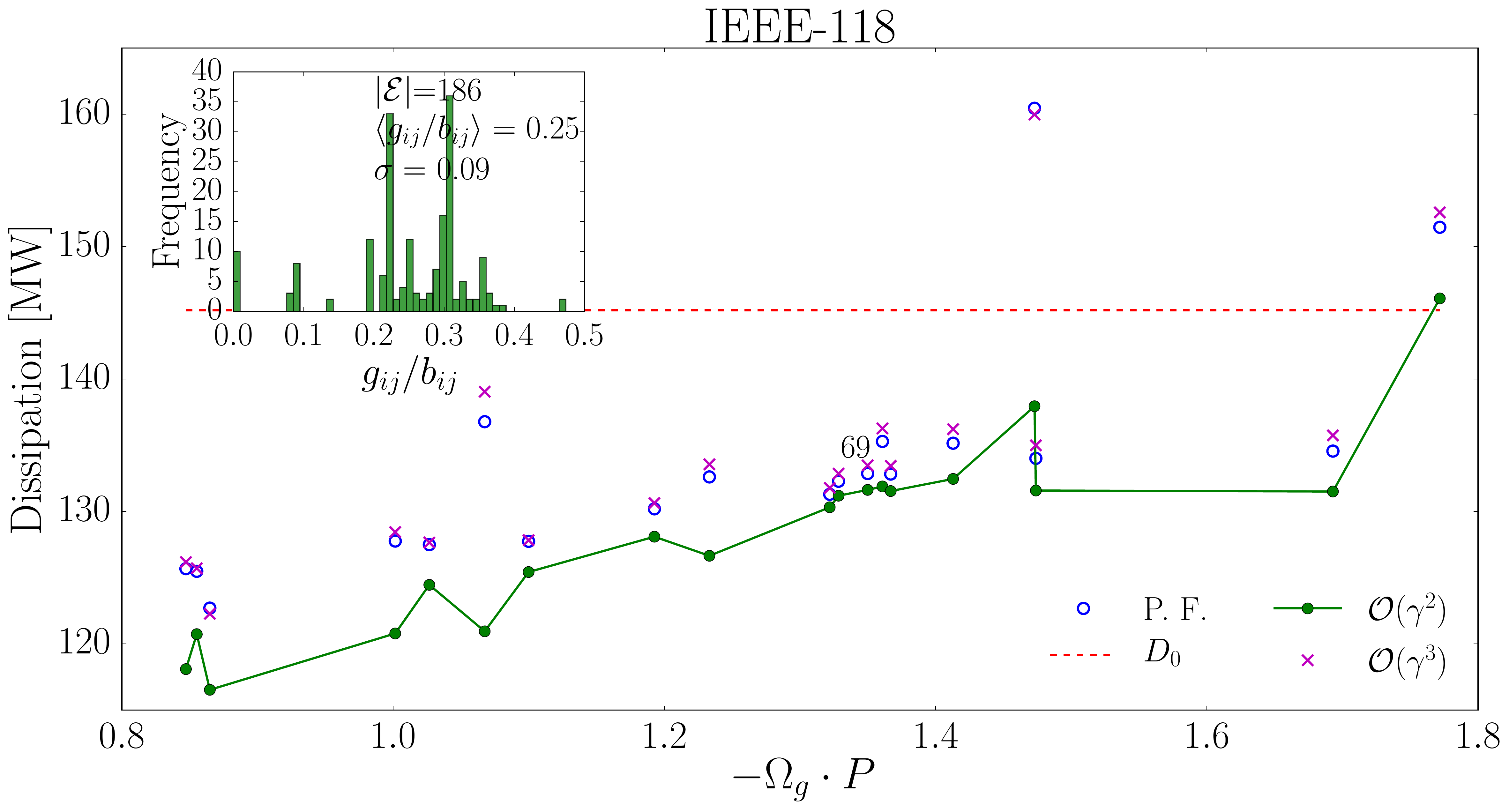}
  \includegraphics[width=0.49\textwidth]{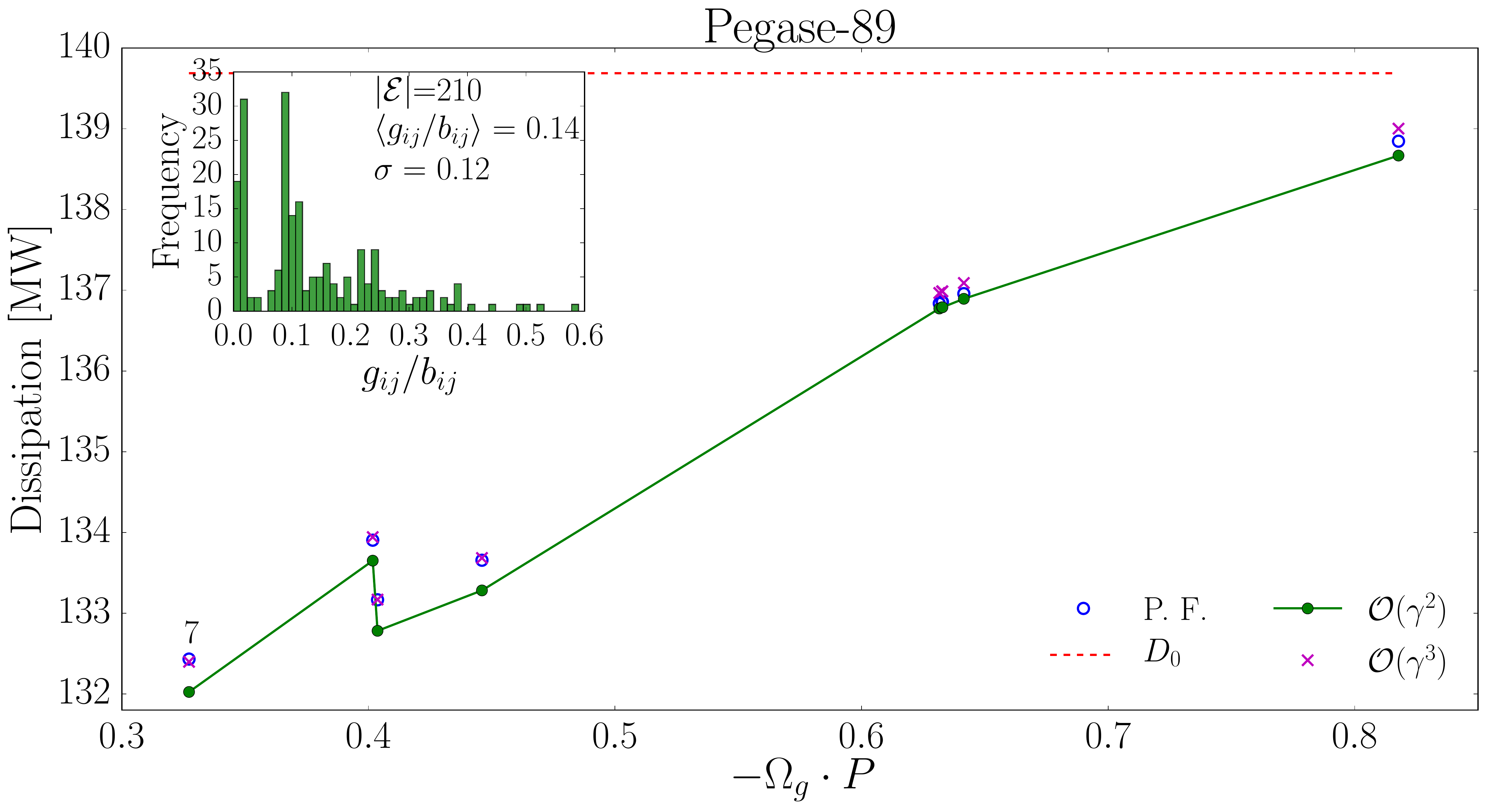}
  \includegraphics[width=0.49\textwidth]{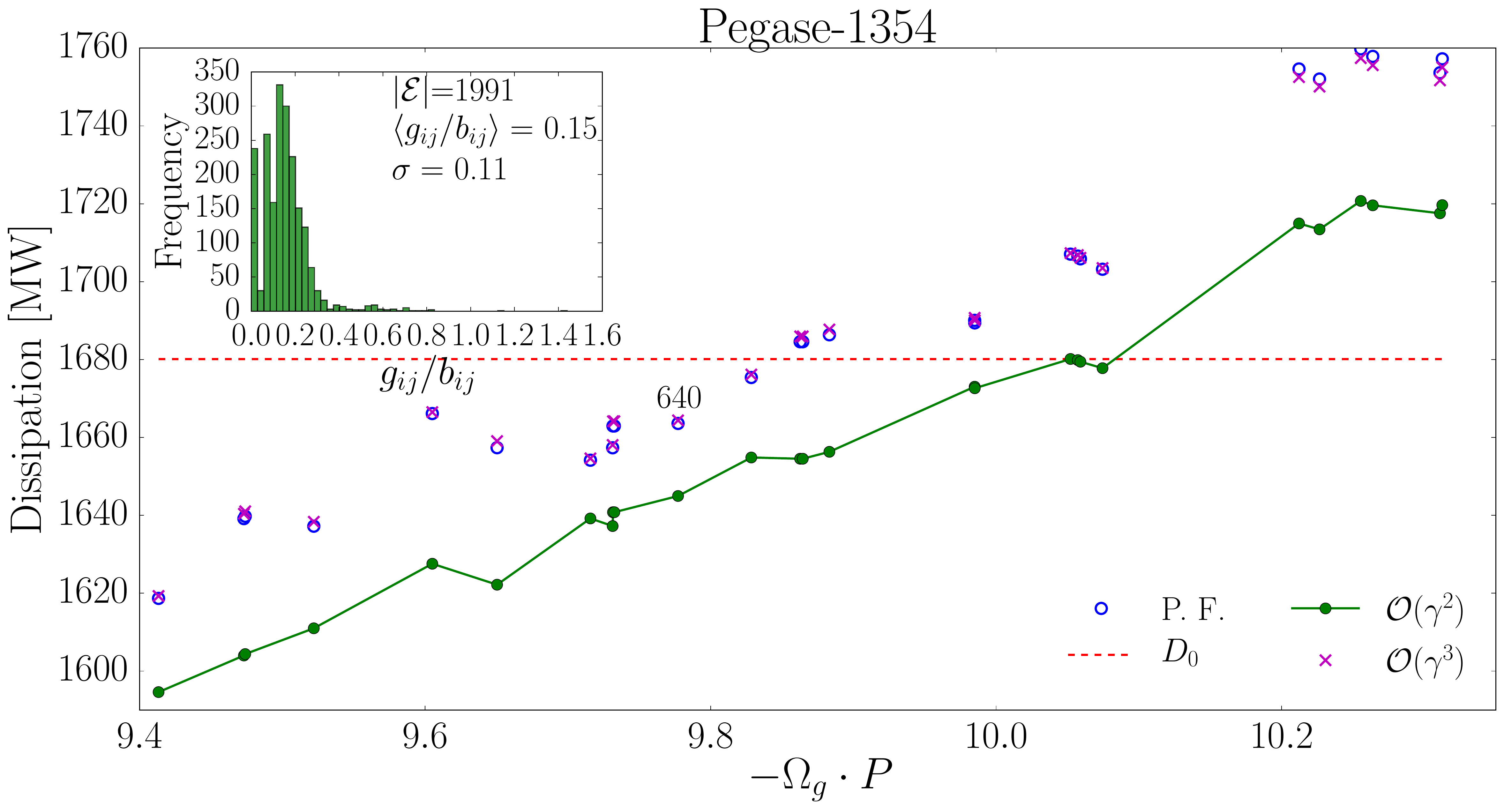}
 \caption{Total dissipation as a function of $-\bm{\Omega_g}^\top \cdot \bm{P}$, 
 for different slack bus choices for the same testcases as in Fig.~\ref{fig:testcases uniform lines}
 (restricted to generators having $P_i>1$GW for the Pegase-1354 testcase). 
 The insets present the distribution of conductance to susceptance ratios extracted from the tabulated values.
 The dashed red line and the green line are $D^{(0)}$ and the expression of Eq.~(\ref{eq:Total dissipation compact}) respectively. 
 Pink crosses are $\mathcal{O}(\gamma^3)$ estimates for the total dissipation [sum of Eq.~(\ref{eq:Total dissipation compact})
 and Eq.~(\ref{eq:Second order term easy})] while blue circles are 
 the Newton Raphson numerical solutions of the full power flow problem.
 The data points annotated by a number indicate the tabulated slack bus generators.}
 \label{fig:testcases tabulated lines}
\end{figure*}

Second, we repeat these simulations for the tabulated line conductance values. In this case the ratio $g_{ij}/b_{ij}$ 
is no longer constant and varies from line to line. The insets of Fig.~\ref{fig:testcases tabulated lines} 
show the distribution of $g_{ij}/b_{ij}$ for all lines in the grid.
Despite varying $g_{ij}/b_{ij}$, Fig.~\ref{fig:testcases tabulated lines} confirms qualitatively that the total dissipation 
is close to being linear in $-\bm{\Omega}_g^\top\cdot\bm{P}$.
Again, tabulated slack buses are not the choice minimizing transmission losses. 
The latter differ by up to $10\%$ depending on the choice of a slack bus.

\subsection{Optimal power dispatch to leading order in $\gamma$}\label{sec:Multiple generation compensation}
We next consider the situation where all generating units contribute to compensating the transmission losses, and investigate
what is the optimal power dispatch that minimizes them. 
We show that up to order $\gamma^2$ the optimal choice is that of a single slack.
We consider $\bm{\delta P}$ of the form
\begin{align}\label{eq:General dispatch}
\nonumber
&\delta P_i\geq0\,, \quad i\in\{\textrm{Gen.}\}\,, \\
&\delta P_i=0\,, \quad i\in\{\textrm{Cons.}\}\,, 
\end{align}
where $\{\textrm{Gen.}\}$ and $\{\textrm{Cons.}\}$ respectively are the sets of generator and consumer indices. 
Fixing $\delta P_i=0$ for consumers is motivated by the fact that we are interested in the optimal power dispatch that delivers 
a fixed amount of power to the consumers -- consumers do not curtail their power demand.
Allowing $\delta P_i<0$ at generator nodes would imply that some generators inject less power 
than the scheduled value $P_i$, a complete rescheduling 
of the power injections that is beyond the scope of this work.
Minimizing the losses amounts to minimizing Eq.~(\ref{eq:Total dissipation dispatch}), which we rewrite using 
Eqs.~(\ref{eq:Product resistance distance power}) and (\ref{eq:General dispatch}) as
\begin{equation}\label{eq:Function to minimize}
D|_{\bm{\delta P}^{(1)}}=\gamma^2\sum_{g \in\{\textrm{Gen}\}}\delta P_g^{(1)}
                    \left[-\bm{\Omega_g}^\top \cdot \bm{P}+\sum_{l\geq2, i}
                    \lambda_l^{-1}P_i{u_i^{(l)}}^2\,\right]\,,
\end{equation}
under the constraint that 
\begin{equation}\label{eq:Constraint}
 \sum_{g\in\{\textrm{Gen.}\}}\gamma\delta P_g^{(1)}=D^{(0)}\,.
\end{equation}
Up to and including the order $\mathcal{O}(\gamma^2)$, the losses, Eq.~(\ref{eq:Function to minimize}), are linear in $\delta P_g^{(1)}$, 
therefore $D|_{\bm{\delta P}^{(1)}}$ is minimal for 
\begin{equation}\label{eq:Solution distributed slack}
 \gamma\delta P_g^{(1)}=\left\{\begin{array}{l}
                  \displaystyle D^{(0)} \quad \textrm{if } g=\bar{g}\,, \quad \quad 
			    \bar{g}\equiv\argmin_{g\in\{\textrm{Gen.}\}}(-\bm{\Omega_g}^\top \cdot \bm{P})\,, \\
                  0 \quad \textrm{otherwise}\,.
                   \end{array}\right.
\end{equation}

\begin{figure*}[!t]
\centering
  \includegraphics[width=0.32\textwidth]{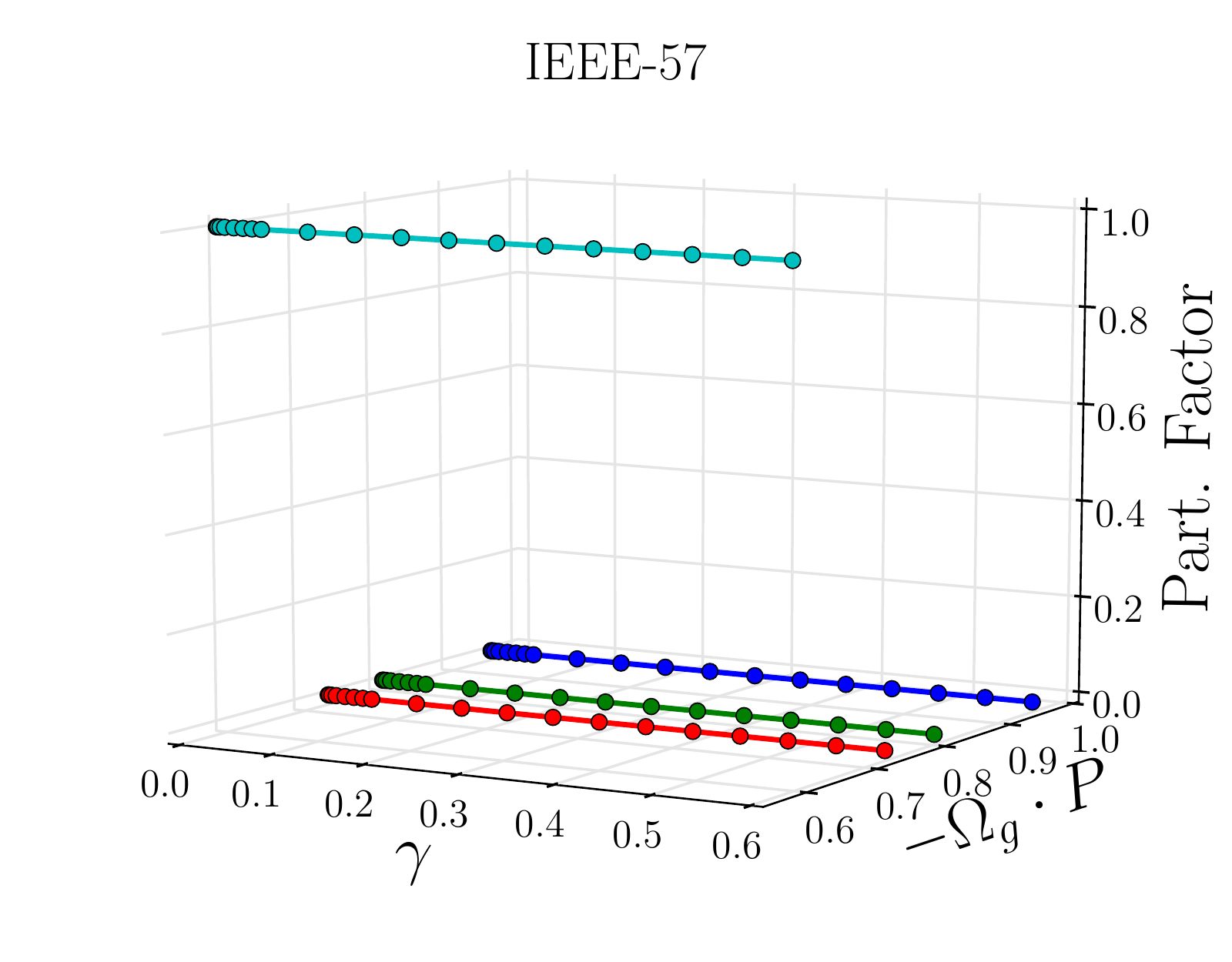}
  \includegraphics[width=0.32\textwidth]{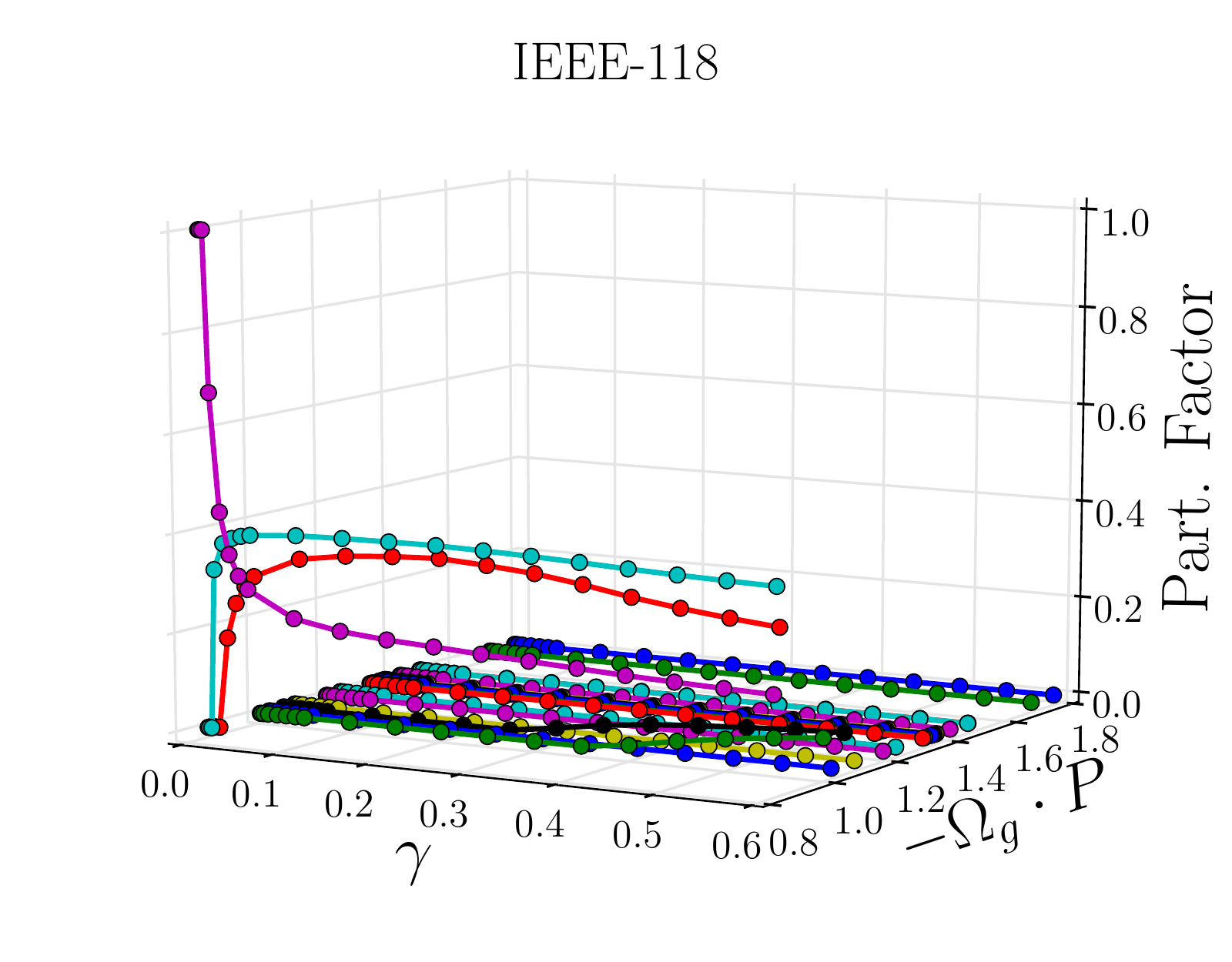}
  \includegraphics[width=0.32\textwidth]{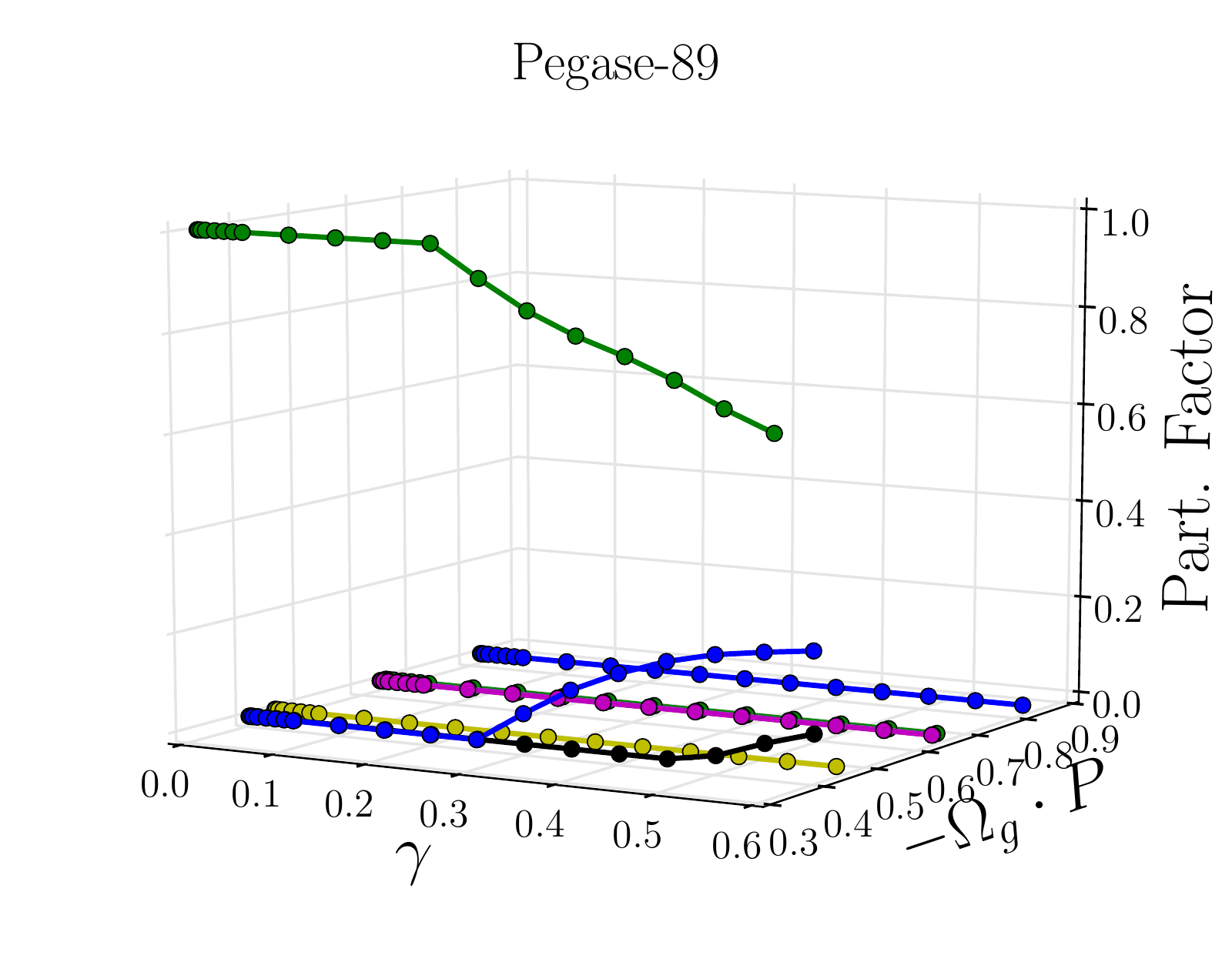}
 \caption{
 Participation factors of the different slack bus candidates obtained by minimizing the 
 approximate total losses [sum of Eqs.~(\ref{eq:Total dissipation compact}) and (\ref{eq:Second order term easy})],
 under the constraints 
Eqs.~(\ref{eq:General dispatch}) and (\ref{eq:Constraint}).
 The minimizations are performed for different values of $\gamma=g_{ij}/b_{ij}\in[0,0.6]$.}
 \label{fig:contribution factors}
\end{figure*}

This shows that the power dispatch scheme that minimizes the total transmission losses to order 
$\mathcal{O}(\gamma^2)$ is the single generator compensation. 
This somehow counterintuitive result is in agreement with the conclusions of Ref.~\cite{Exposito04}.
Eq.~(\ref{eq:Function to minimize}) rephrases these results by connecting the slack bus choice that minimizes losses 
to the resistance distance. This distance measure accounts for the multiple paths between the slack generator and any 
other node of the network, weighted by the load of the lines in the lossless case.

In the next section we investigate the impact that truncating the calculation to order $\mathcal{O}(\gamma^2)$
has on the solution of the transmission loss minimization problem.

\subsection{Next to leading order contributions and implications for the optimal power flow}\label{sec:NLO}

So far we have considered terms up to order $\mathcal{O}(\gamma^2)$ in the total dissipation. The next to leading order contribution 
is proportional to $\gamma^3$ and is given by
\begin{multline}\label{eq:Next order Dissipation}
\displaystyle \gamma^3\sum_{\langle i,j\rangle}b_{ij}V_iV_j\left[
2\sin(\theta_i^{(0)}-\theta_j^{(0)})(\delta\theta_i^{(2)}-\delta\theta_j^{(2)})+\right. \\
\left.\cos(\theta_i^{(0)}-\theta_j^{(0)})(\delta\theta_i^{(1)}-\delta\theta_j^{(1)})^2
\right]\,.
\end{multline}

Eq.~(\ref{eq:Next order Dissipation}) consists of two contributions: the first which is linear in the second order expansion of 
the angle variations $\bm{\delta\theta}^{(2)}$, and the second which is quadratic in the first order contribution of the angle 
variations $\bm{\delta\theta}^{(1)}$. 
Since the first term in the right-hand side of Eq.~(\ref{eq:Next order Dissipation}) is proportional to the sine while the second term
to the cosine of the phase differences, for weakly loaded networks we expect the contribution of the latter to be the dominant one.
While computing $\bm{\delta\theta}^{(2)}$ is quite involved, 
one can easily evaluate the second term using Eq.~(\ref{eq:Delta theta}).
One obtains
\begin{multline}\label{eq:Second order term easy}
\displaystyle \gamma^3\sum_{\langle i,j\rangle}b_{ij}V_iV_j\cos(\theta_i^{(0)}-\theta_j^{(0)})
(\delta\theta_i^{(1)}-\delta\theta_j^{(1)})^2 = \\
 \gamma^3\sum_{\langle i,j\rangle}b_{ij}V_iV_j\cos(\theta_i^{(0)}-\theta_j^{(0)})\\
 \cdot\left(\sum_{l\geq2}\lambda_l^{-1}\left(u_i^{(l)}-u_j^{(l)}\right)\left[{{\bm u}^{(l)}}^\top\cdot(\bm{\delta P}^{(1)}-{\bm v})\right]\right)^2\,,
\end{multline}
which is a positive quantity if all phase differences $|\theta_i^{(0)}-\theta_j^{(0)}|\leq\pi/2$.
This implies that for normally loaded networks, this next to leading order contribution always increases the losses
with respect to the lower order estimate, Eq.~(\ref{eq:Total dissipation compact}).

Computing the contribution of Eq.~(\ref{eq:Second order term easy}) to the total dissipation
requires the knowledge of the lossless power flow solution only.
In the case of the single slack bus compensation scheme, one takes for $\gamma\bm{\delta P}^{(1)}$ the vector with components 
$\gamma\delta P_i^{(1)}=D^{(0)}\delta_{i,g}$ to evaluate Eq.~(\ref{eq:Second order term easy}).
Pink crosses in Figures \ref{fig:testcases uniform lines} and \ref{fig:testcases tabulated lines} present the numerical evaluations of 
Eq.~(\ref{eq:Second order term easy}). 
Despite, the fact that considering only Eq.~(\ref{eq:Second order term easy}) for the total dissipation
to order $\mathcal{O}(\gamma^3)$ is not a controlled approximation, the numerics confirm that it is a quantitatively very accurate one 
for normally loaded networks.
This justifies a posteriori the assumption that Eq.~(\ref{eq:Second order term easy}) is the dominant contribution of Eq.~(\ref{eq:Next order Dissipation}).

Our calculation shows that to order $\mathcal{O}(\gamma^2)$, 
the total losses are minimized by choosing as unique slack the generator having the minimal
projected resistance distance $-\bm{\Omega}_g\cdot\bm{P}$.
It is natural to ask whether this conclusion holds beyond $\mathcal{O}(\gamma^2)$.
To answer this question we numerically minimize the total losses given by the sum of terms proportional to $\gamma^2$ 
and $\gamma^3$, Eqs.~(\ref{eq:Dissipation expansion2}) and (\ref{eq:Second order term easy}), under the constraints 
Eqs.~(\ref{eq:General dispatch}) and (\ref{eq:Constraint}).
This constrained minimization problem is quadratic in the variables $\{\delta P_i^{(1)}\}$.
The outcome of the minimization yields the power dispatch $\bm{\delta P}_\star^{(1)}$ that minimizes the losses. 

The components of $\bm{\delta P}_\star^{(1)}/D^{(0)}$, which correspond to the relative participation factors of the different generators,
are presented in Figure~\ref{fig:contribution factors} for different testcases and different ratios of $g_{ij}/b_{ij}=\gamma$.
For the testcase IEEE-57 the first order prediction is very robust. For all values of $\gamma\in[0,0.6]$ the losses are minimal 
if only the generator having the lowest $-\bm{\Omega}_g\cdot\bm{P}$ increases its production. Out of the four generator slack candidates,
the participation factors are all identical to zero except for one generator for which it is maximal.

In contrast, for the IEEE-118 testcase we find that a combination of three generators increasing their
injection is the power dispatch which minimizes losses already for moderate values of $\gamma$. 
For very low values of $\gamma$, the generator having the lowest
$-\bm{\Omega}_g\cdot\bm{P}$ has a participation factor equal to one while it is equal to zero for all other generators,
as predicted by Eq.~(\ref{eq:Solution distributed slack}).
However, for $\gamma\gtrsim0.02$ the three generators having the smallest $-\bm{\Omega}_g\cdot\bm{P}$ values roughly all contribute for
$1/3$ of the losses. The lower order prediction that single generator compensation is optimal stops holding already at such low values of the expansion parameter 
because three generators have almost degenerate values of $-\bm{\Omega}_g\cdot\bm{P}$ as can be seen in Fig.~\ref{fig:testcases uniform lines}
(second row).

Finally, the Pegase-89 testcase displays a similar behavior, where single 
generator compensation is optimal as long as $\gamma\lesssim0.25$ at which point the injection of two other generators picks up.
We conclude that the single slack choice is the optimal choice except when the two (or more) smallest values of $-\bm{\Omega}^\top\cdot\bm{P}$
are almost the same, for finite values of $\gamma$.

\section{Conclusion}\label{sec:Conclusion}
We have investigated the dependence of transmission losses on the slack bus selection.
Our analytical approach, valid for transmission lines having small, homogeneous ratios $r/x=\gamma$ indicates that, generically, transmission losses are minimal if a 
unique \textit{slack bus} injects the additional power required to compensate for transmission losses.
To leading order in $r/x$,  we show that the optimal slack bus choice is the generator for which the 
graph theoretical metric $-\bm{\Omega}_g^\top\cdot\bm{P}$ is minimal. This is a computationally inexpensive quantity to evaluate.
It is the average of the resistance distance separating the slack bus candidate from all other nodes of the network, weighted by the power injections of the lossless 
problem.

For larger values of $r/x$, we show that the optimal choice is a distributed slack if several
generators have almost the same, smallest values of $-\bm{\Omega}_g^\top\cdot\bm{P}$, and that this effect is of order $\mathcal{O}(\gamma^3)$.
The metric we propose could be used to provide a hot start to more sophisticated optimal power flow algorithms
since it specifies the group of generators which are most relevant for transmission losses. 
While the effect found here is rather small, with transmission losses reduced by $\sim10\%$ with an optimal
slack bus choice, we think that picking the optimal slack bus choice may be more crucial in future AC networks
with larger shares of renewables, and more heavily loaded networks than the standard testcases investigated in 
this manuscript.

Finally, we predict that the resistance distance is the relevant metric for many other electric networks problems 
dealing with identifying critical nodes or lines, and which involve inverting a graph Laplacian.

\section*{Acknowledgment}
T.~C. thanks F.~D\"{o}rfler for useful discussions. This work was supported by the Swiss National Science Foundation
under an AP Energy Grant.

\bibliographystyle{IEEEtran}
\bibliography{IEEEabrv,bibliography}

\begin{IEEEbiographynophoto}{Tommaso Coletta}
Received his M.Sc. degree in physics and his Ph.D. degree in theoretical physics from the 
Ecole Polytechnique F\'ed\'erale de Lausanne (EPFL), Lausanne, Switzerland in 2009 and 2013 respectively.
He has been a Postdoctoral researcher at the Chair of Condensed Matter Theory at the Institute of Theoretical Physics of EPFL.
Since 2014 he is a Postdoctoral researcher at the engineering department of the University of 
Applied Sciences of Western Switzerland, Sion, Switzerland working on complex networks and power systems.
\end{IEEEbiographynophoto}

\begin{IEEEbiographynophoto}{Philippe Jacquod}
Philippe Jacquod received the Diplom degree in theoretical physics from the ETHZ, Z\"{u}rich, Switzerland, in 1992, and the PhD degree in 
natural sciences from the University of Neuch\^{a}tel, Neuch\^{a}tel, Switzerland, in 1997. He is a professor with the engineering department,
University of Applied Sciences of Western Switzerland, Sion, Switzerland. From 2003 to 2005 he was an assistant professor with the 
theoretical physics department, University of Geneva, Geneva, Switzerland and from 2005 to 2013 he was associate, then full professor 
with the physics department, University of Arizona, Tucson, USA. Currently, his main research topics is in power systems and how they
will evolve as the energy transition unfolds. He is co-organizing an international conference series on that topics. He has published 100 
papers in international journals, books and conference proceedings. 
\end{IEEEbiographynophoto}

\end{document}